\documentclass[useAMS,usenatbib]{mn2e}
\usepackage{natbib}
\usepackage{graphicx}
\usepackage{times}
\usepackage{amssymb, amsmath}
\usepackage{epstopdf}
\usepackage{color}
\usepackage[section]{placeins}
\usepackage[english]{babel}
\usepackage{lineno}
\usepackage{url}
\usepackage{ulem}
\usepackage[colorlinks,linkcolor={blue},citecolor={blue},urlcolor={blue}]{hyperref}
\def\mpchi{\,h^{-1}{\rm {Mpc}}}

\def\kms{\,{\rm {km\, s^{-1}}}}
\def\msun{\,h^{-1}{\rm M_{\sun}}}

\def\r{\mathbf{r}}

\def\apj{ApJ}

\def\apjl{ApJL}

\def\mnras{MNRAS}

\def\prd{PhRvD}

\begin{document}

\title[Modelling the Redshift-Space 3PCF]{Modelling The Redshift-Space Three-Point Correlation Function in SDSS-III}

\author[Guo et al.]{\parbox{\textwidth}{
Hong Guo$^{1}$\thanks{E-mail: hong.guo@utah.edu}, Zheng Zheng$^{1}$, Y.~P.
Jing$^{2,3}$, Idit Zehavi$^{4}$, Cheng Li$^{5}$, David H. Weinberg$^{6,7}$,
Ramin A. Skibba$^{8}$, Robert C. Nichol$^{9}$, Graziano Rossi$^{10,11}$,
Cristiano G. Sabiu$^{12}$, Donald P. Schneider$^{13,14}$, Cameron K.
McBride$^{15}$}
\vspace*{6pt} \\
$^{1}$ Department of Physics and Astronomy, University of Utah, UT 84112, USA\\
$^{2}$ Center for Astronomy and Astrophysics, Department of Physics and
Astronomy, Shanghai Jiao Tong University,
Shanghai 200240, China \\
$^{3}$ IFSA Collaborative Innovation Center, Shanghai Jiao Tong University, Shanghai 200240, China\\
$^{4}$ Department of Astronomy, Case Western Reserve University, OH 44106, USA\\
$^{5}$ Partner Group of the Max Planck Institute for Astrophysics and Key
Laboratory for Research in Galaxies and
Cosmology \\
  ~~~of Chinese Academy of Sciences, Shanghai Astronomical Observatory, Nandan Road 80, Shanghai 200030, China\\
$^{6}$ Department of Astronomy, Ohio State University, Columbus, OH 43210, USA\\
$^{7}$ Center for Cosmology and Astro-Particle Physics, Ohio State University, Columbus, OH 43210, USA\\
$^{8}$ Center for Astrophysics and Space Sciences, University of California,
9500 Gilman Drive, San Diego, CA 92093,
USA\\
$^{9}$ Institute of Cosmology \& Gravitation, Dennis Sciama Building,
University of Portsmouth, Portsmouth, PO1 3FX,
UK\\
$^{10}$ Department of Astronomy and Space Science, Sejong University, Seoul, 143-747, Korea\\
$^{11}$ CEA, Centre de Saclay, Irfu/SPP, F-91191 Gif-sur-Yvette, France\\
$^{12}$ Korea Institute for Advanced Study, Dongdaemun-gu, Seoul 130-722,
Korea\\
$^{13}$ Department of Astronomy and Astrophysics, The Pennsylvania State
University, University Park, PA 16802\\
$^{14}$ Institute for Gravitation and the Cosmos, The Pennsylvania State
University, University Park, PA 16802\\
$^{15}$ Harvard-Smithsonian Center for Astrophysics, 60 Garden St.,
Cambridge, MA 02138, USA}

\maketitle

\begin{abstract}
We present the measurements of the redshift-space three-point correlation
function (3PCF) for $z\sim 0.5$ luminous red galaxies of the CMASS sample in
the Sloan Digital Sky Survey-III Baryon Oscillation Spectroscopic Survey Data
Release 11. The 3PCF measurements are interpreted within the halo occupation
distribution (HOD) framework using high-resolution N-body simulations, and
the model successfully reproduces the 3PCF on scales larger than $1\mpchi$.
As with the case for the redshift-space two-point correlation functions, we
find that the redshift-space 3PCF measurements also favour the inclusion of
galaxy velocity bias in the model. In particular, the central galaxy in a
halo is on average in motion with respect to the core of the halo. We discuss
the potential of the small-scale 3PCF to tighten the constraints on the
relation between galaxies and dark matter haloes and on the phase-space
distribution of galaxies.
\end{abstract}


\section{Introduction}
Contemporary galaxy redshift surveys enable the large-scale distribution of
galaxies to be accurately mapped in redshift space. Compared to the
real-space distribution, that in redshift space is distorted as a result of
galaxy peculiar velocities, which is generally referred to as redshift space
distortions (RSD). The RSD effects encode information about the kinematics of
galaxies inside dark matter haloes and the growth rate of cosmic structure.

The clustering of galaxies in redshift space has been extensively studied
using the two-point correlation functions (2PCFs)
\citep{Zehavi02,Zehavi05b,Zehavi11,Wang04,Wang07,Li06,Coil06,Skibba09b,Li12,Guo13}.
The widely-used halo occupation distribution (HOD) modelling of the galaxy
2PCFs provides an opportunity to understand the connection between galaxies
and their host dark matter haloes
\citep{Jing98,Peacock00,Scoccimarro01,Berlind02,Zheng05,Zheng09,Miyatake13,Guo14a}.
The observationally constrained relation between galaxies and dark matter
haloes provides insight also about galaxy formation and cosmology. Recently,
\cite{Guo15} (hereafter G15) used the luminous red galaxies (LRGs) in the
Sloan Digital Sky Survey-III \citep[SDSS-III;][]{Eisenstein11} Baryon
Oscillation Spectroscopic Survey \citep[BOSS;][]{Dawson13} to model the
redshift-space 2PCFs and found differences between galaxy and dark matter
velocity distributions inside haloes, an effect denoted as galaxy velocity
bias (see also \citealt{Reid14}).

Higher-order statistics, e.g. the three-point correlation function (3PCF),
aid in tightening the constraints on HOD parameters and in breaking the
degeneracy among parameters \citep{Kulkarni07,Smith08}. The 3PCF,
$\zeta(r_1,r_2,r_3)$, describes the probability of finding galaxy triplets
with the separations in between as $r_1$, $r_2$ and $r_3$ (see
\citealt{Bernardeau02} for a review). A non-zero 3PCF naturally arises
because of the non-Gaussianity generated during the nonlinear evolution of
the density fluctuations, even if primordial fluctuations were perfectly
Gaussian. The 3PCF is commonly used to break the degeneracy between the
galaxy bias and the amplitude of the matter density fluctuation, and
therefore constrains cosmological parameters
\citep{Gaztanaga94,Jing04,Gaztanaga05,Zheng04,Pan05,Guo09b,McBride11b,Marin13,Guo14b}.

In this Letter, we measure the redshift-space 3PCF for the same sample of LRGs
at redshift $z{\sim}0.5$ as in G15 and interpret the 3PCF measurements within
the HOD framework. In particular, we perform HOD modelling of both the 2PCFs
and 3PCF and investigate the additional constraining power from the 3PCF on
the HOD parameters, including the galaxy velocity bias.

In Section~\ref{sec:data}, we briefly describe the galaxy sample, the 3PCF
measurements, and the modelling method. We present our modelling results in
Section~\ref{sec:results} and conclude in Section~\ref{sec:discussion}.
Throughout this Letter we adopt a spatially flat $\Lambda$CDM cosmology with a
matter density parameter $\Omega_m=0.27$, $\sigma_8=0.82$ and a Hubble
constant $H_0=100h$\,kms$^{-1}$Mpc$^{-1}$ with $h=0.7$.

\section{Data and Measurements}
\label{sec:data} In this Letter, we use the same volume-limited LRG sample as
in G15 ($i$-band absolute magnitude $M_i<-21.6$ and $0.48<z<0.55$), selected
from SDSS-III BOSS CMASS galaxies \citep{Eisenstein11,Bolton12}. Following
\citet{Guo14b}, the 3PCF is calculated using the estimator of
\cite{Szapudi98}. The triangles are represented in the parametrization of
$(r_1,r_2,\theta)$, with $r_2\ge r_1$ and $\theta$ being the angle between
$\r_1$ and $\r_2$. We use linear binning schemes for $r_1$, $r_2$ and
$\theta$, with $\Delta r_1=\Delta r_2=2\mpchi$, and $\Delta\theta=0.1\pi$.

To correct for the fibre collision effect in the redshift-space 3PCF,
$\zeta(r_1,r_2,\theta)$, we assign the redshift of each fibre-collided galaxy
from its nearest neighbour \citep{Guo12,Guo14b}. To further reduce any
residual effect on scales slightly larger than the projected fibre collision
scale in SDSS-III (${\sim}0.5\mpchi$) \citep{Gunn06,Dawson13,Smee13}, we only
consider the triangle configurations of $r_2=2r_1$ and $r_1\ge1\mpchi$, which
will ensure that all sides of the triangle are larger than $1\mpchi$. We
measure the redshift-space 3PCF for six $r_1$ bins, centred at $2\mpchi$,
$4\mpchi$, ..., $12\mpchi$, with each bin having 10 triangle configurations
of different values of $\theta$.

\begin{figure}
\includegraphics[width=0.55\textwidth]{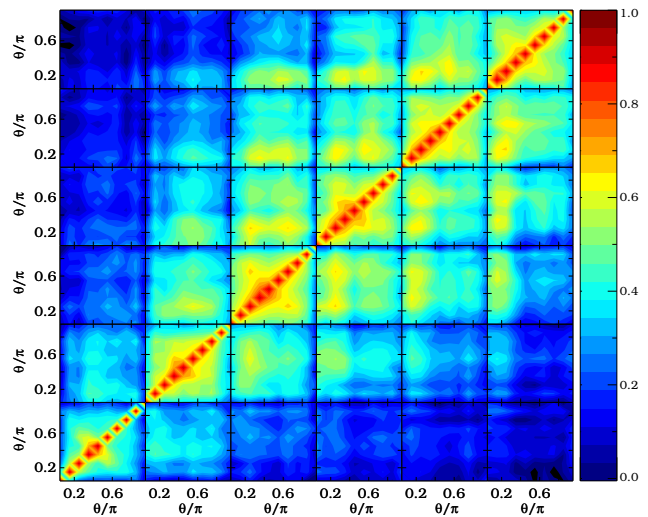}
\caption{Normalized covariance matrix of the 3PCF. A galaxy triplet is
represented by two sides of the triangle, $r_1$ and $r_2=2 r_1$, and the
angle $\theta$ between them. From left to right and bottom to top, the plot
shows the normalized covariance matrix for the measurements
$\zeta(r_1,r_2,\theta)$ in 10 different
$\theta$ bins for $r_1$ centred at 2, 4, 6, 8, 10, and 12$\mpchi$.
\label{fig:cov}
}
\end{figure}
\begin{figure*}
\includegraphics[width=1.0\textwidth]{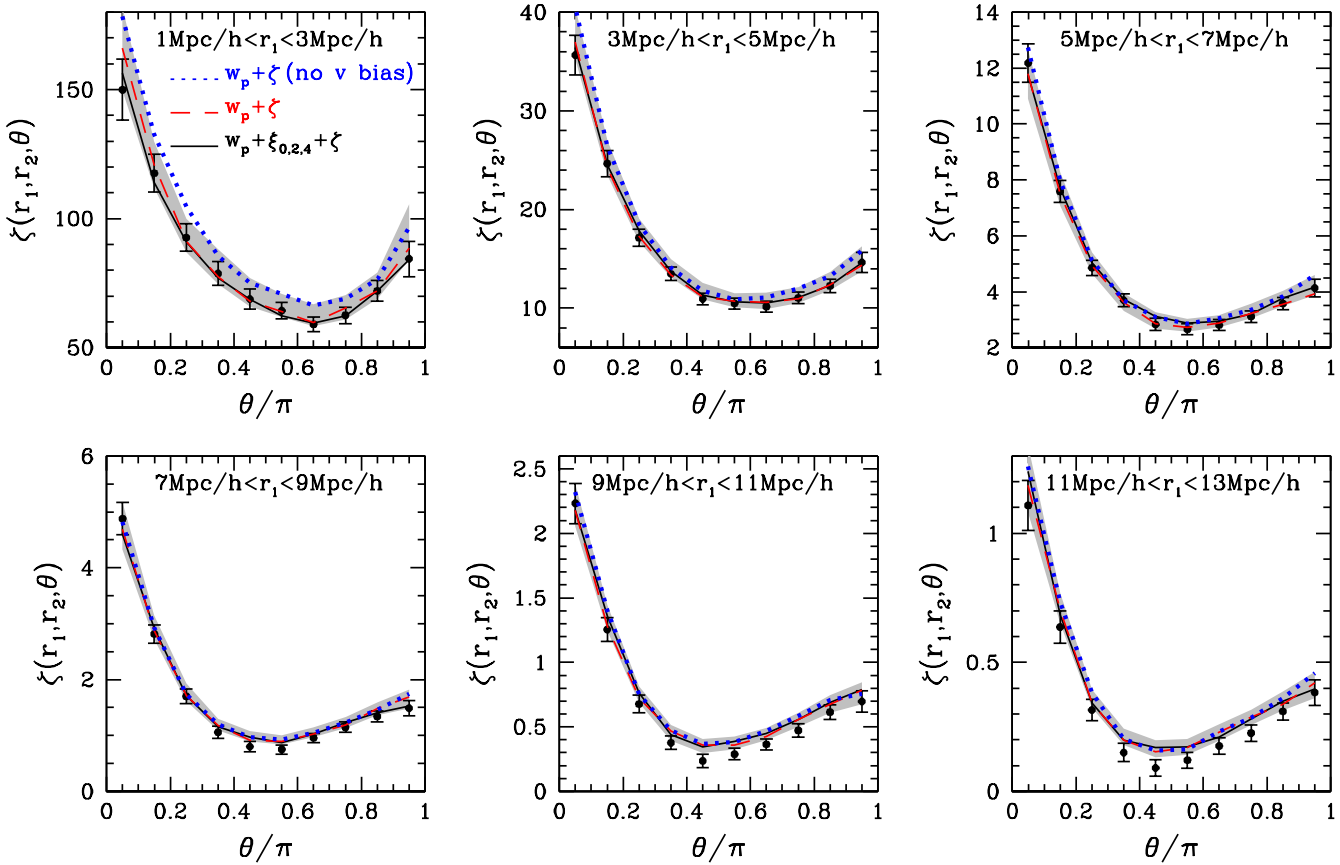}
\caption{Comparison of the 3PCF measurements (circles with error bars) for
BOSS CMASS galaxies to the best-fitting models from HOD fitting to
$w_p+\zeta$ (with and without velocity bias) and $w_p+\xi_{0,2,4}+\zeta$. The
shaded areas correspond to the 1$\sigma$ uncertainties in the best-fitting
model from fitting $w_p+\xi_{0,2,4}+\zeta$. Different panels display the
measurements $\zeta(r_1,r_2,\theta)$ as a function of $\theta$ in different
$r_1$ bins ($r_2=2r_1$). } \label{fig:zeta}
\end{figure*}
For HOD modelling, we also adopt the 2PCF measurements from G15 and perform a
joint fit with the 3PCF, which allows a study of the information content
related to the HOD constraints from the 3PCF. The 2PCF measurements include
the projected 2PCF ($w_p$), the monopole ($\xi_0$), quadrupole ($\xi_2$) and
hexadecapole moments ($\xi_4$) of the redshift-space 2PCF from $0.13$ to
$51.5\mpchi$ (see G15).  The full covariance matrix (including the
cross-correlation between the 2PCF and 3PCF measurements) is estimated from
$N_{jk}=403$ jackknife subsamples (G15). To reduce the effect of noise in the
covariance matrix, we apply the method of \cite{Gaztanaga05b} and only retain
the modes that have eigenvalues $\lambda^2>\sqrt{2/N_{jk}}$.
Figure~\ref{fig:cov} displays the covariance matrix of the 3PCF. In each
$r_1$ bin, the measurements of different $\theta$ bins are strongly
correlated with each other. The correlation between different $r_1$ bins is
stronger for larger $r_1$, implying better constraints to the model from the
measurements at smaller scales.

Following G15, we adopt the five-parameter HOD model of \cite*{Zheng07} and
include two additional velocity bias parameters, $\alpha_c$ and $\alpha_s$,
for central and satellite galaxies, respectively. To model the redshift-space
2PCF, we use the simulation-based model of Zheng \& Guo (in preparation; see
also G15), which is equivalent to populating haloes in the simulations with
galaxies for a given HOD prescription. We use the $z=0.53$ dark matter halo
catalogs from the MultiDark simulation \citep[see details in][]{Prada12}. The
haloes are defined using the spherical overdensity (SO) algorithm, with a
mean density ${\sim}237$ times that of the background universe at $z=0.53$.
The halo centre is defined as the position of the dark matter particle with
the minimal potential\footnote{There can be chances that the potential
minimum is found in a sub-halo passing close to the halo centre. Even in such
a case, assigning the central galaxy to the position of the potential minimum
would not affect our modelling result, since the offset from the centre is
likely to be smaller than the smallest scales probed by our clustering
measurements. In fact, the existence of the central velocity bias itself
points to an offset between the central galaxy and halo centre, which is
again small compared to the scales we probe (see G15 for more discussions on
the expectation and comparison to other observations).}. We choose the bulk
velocity of the inner $25$ percent halo particles around the potential
minimum as the halo (core) velocity, and the velocity bias is parametrized
in this frame, as in G15.

We place the central galaxies at the halo centres, with their velocities the
same as the halo velocities. For satellite galaxies, we assign the positions
and velocities of randomly-selected dark matter particles. We then add the
velocity bias effect. The central galaxy velocity bias is parametrized as an
additional Gaussian component with zero mean and a standard deviation of
$\alpha_c\sigma_v$, where $\sigma_v$ is the line-of-sight (LOS) velocity
dispersion of the dark matter particles in each halo. For the satellite
velocity bias, the relative LOS velocity of a satellite galaxy to the halo
core is scaled by $\alpha_s$ (as detailed in G15), therefore the 1D satellite
galaxy velocity dispersion $\sigma_s$ is a factor of $\alpha_s$ times that of
the dark matter particles, $\sigma_s=\alpha_s\sigma_v$.

To model the 2PCFs and 3PCFs, we apply a Markov chain Monte Carlo method to
explore the HOD parameter space. For each set of HOD parameters, the 2PCFs
are calculated using the simulation-based model (G15), while the 3PCFs are
directly measured from the mock catalogs. The value of $\chi^2$ is the sum of
the contribution from the correlation functions and that from the number
density of the sample.

\section{Results}\label{sec:results}
\begin{figure*}
\includegraphics[width=0.7\textwidth]{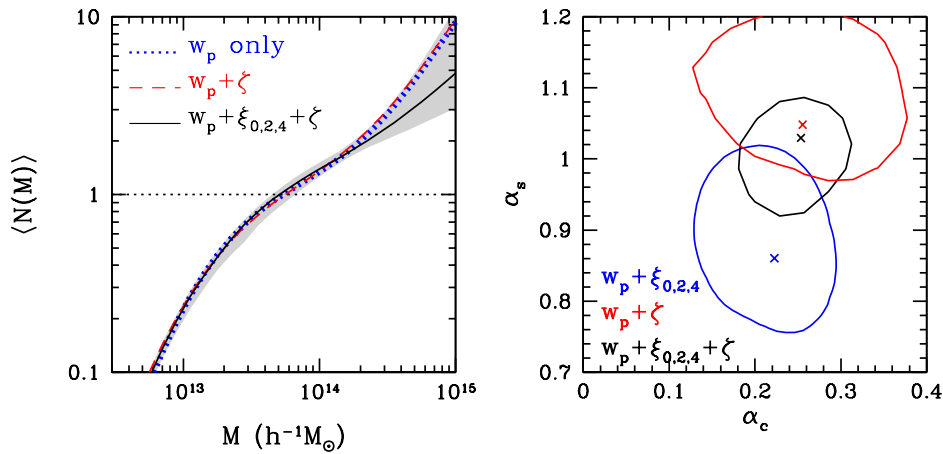}\caption{Left: Mean occupation functions of galaxies from different
models. The shaded area shows the $1\sigma$ distribution around the
best-fitting model from jointly fitting $w_p$, $\xi_{0,2,4}$ and $\zeta$.
Right: Constraints to the galaxy velocity bias parameters from fitting
various combinations of 2PCFs and 3PCF (see the text). The contours show the
95 per cent confidence levels. The crosses correspond to the parameters from
the best-fitting models.} \label{fig:hod}
\end{figure*}
We explore the constraining power of the 3PCFs on the HOD parameters and the
need for including velocity bias parameters by considering three cases. We
include $w_p$ in each case, given its common use in constraining the HOD and
its effectiveness in limiting the parameter space. First, we perform a joint
fit of $w_p$ and the 3PCF $\zeta$ with the 5-parameter HOD model without
velocity bias (i.e. fixing $\alpha_c=0$ and $\alpha_s=1$). This case is
denoted as $w_p+\zeta$ (no $v$ bias). Then, we add $\alpha_c$ and $\alpha_s$
as free parameters in the fitting (7 parameters in total), which is denoted
as $w_p+\zeta$. The two cases would indicate whether velocity bias is favoured
by the data. Finally, we jointly fit all the 2PCF and 3PCF measurements
(denoted as $w_p+\xi_{0,2,4}+\zeta$) with the 7-parameter model to obtain the
HOD constraints that explain all our measurements.

The best-fitting model predictions of the 3PCFs for the case of $w_p+\zeta$
(no $v$ bias) are shown as dotted blue curves in Figure~\ref{fig:zeta}. The
best-fitting $\chi^2$ is 59.6, for 70 degrees of freedom (d.o.f.=70, from 14
$w_p$ points, 60 $\zeta$ points, one number density point, and 5 parameters).
Therefore, the fit without velocity bias is acceptable. Comparing the bestfit
to the $\zeta$ data points, we notice that on small scales ($r_1<5\mpchi$)
the model overpredicts $\zeta$ over all ranges of $\theta$, and on large
scales it overpredicts $\zeta$ for configurations of nearly degenerate
triangles ($\theta\sim 0$ or $\theta\sim \pi$). Even though we cannot judge
the fit by eye, given that the data points are correlated, the comparison
provides hints to the potential improvements that can be achieved in the
modelling. In particular, the small $\chi^2$ could indicate that the
covariance matrix is overestimated, and the above differences would signal an
inadequacy in the model.

We then introduce the velocity bias parameters $\alpha_c$ and $\alpha_s$ in
the modelling. With this 7-parameter model, the best-fitting $\zeta$ is shown
as the dashed red curve in each panel of Figure~\ref{fig:zeta} (labelled as
$w_p$+$\zeta$). Compared to the case without velocity bias, the fit clearly
improves on both small and large scales in a way as expected from the above
discussion. The best-fitting $\chi^2$ becomes 46.4 for 68 degrees of freedom.
Putting aside the possibility of overestimation of the covariance matrix, we
find that both models with and without velocity bias can fit the data well.
Then the question is whether adding the velocity bias is preferred. The
Akaike information criterion (AIC; \citealt{Akaike74}) provides a quantified
way of comparing different models. It penalizes models with more free
parameters, $\rm{AIC}=\chi^2+2N_p$ with $N_p$ the number of free parameters.
To be more conservative, we adopt the AIC with correction (AICc;
\citealt{Konishi07}), $\rm{AICc}=\rm{AIC}+2N_p(N_p+1)/(N_d-N_p-1)$ with $N_d$
the number of data points, which gives more penalty than $\rm{AIC}$ to the
model with more parameters. We find the $\rm{AICc}$ value to be 70.5 and 62.1
for the models without and with velocity bias, respectively. As a
consequence, the model without velocity bias is $\exp((62.1-70.5)/2)=0.015$
times as probable as the model with velocity bias. Therefore, the model with
velocity bias is preferred by the data, even though it includes two more free
parameters. The possibility that the covariance matrix is overestimated would
strengthen this conclusion. The result also implies that the redshift-space
3PCFs can provide constraints on the velocity bias parameters, which is a
point we discuss later.

Finally, using the 7-parameter model we perform a joint fit to $w_p$,
$\xi_{0,2,4}$, and $\zeta$. The best-fitting models are shown as the solid
black curves in Figure~\ref{fig:zeta}, which represents the 3PCF measurements
remarkably well on all scales. The overall best-fitting $\chi^2/\rm{dof}$
value, to all measurements, is $97.4/110$. There is a slight over-prediction
for triangle configurations of $\theta\sim0.4\pi$ on scales of $r_1>7\mpchi$.
According to the covariance matrix in Figure~\ref{fig:cov}, the measurements
for these triangle configurations are strongly correlated among the
neighbouring $\theta$ bins, which implies that the fit cannot be simply judged
by $\chi^2$-by-eye. The shaded regions are the $1\sigma$ uncertainties around
this best-fit model. On large scales ($r_1>7\mpchi$), all three models are
consistent within the $1\sigma$ uncertainties, with the $w_p+\zeta$ (no $v$
bias) model showing the largest deviation. The three models are more readily
separate on smaller scales, where the contribution of the one-halo term (i.e.
the contribution from intra-halo galaxy triplets) becomes important. This
result indicates that the small-scale 3PCF measurements are important in
discriminating the models and in constraining the galaxy distribution within
haloes.

Figure~\ref{fig:hod} shows the constraints on the HOD from different models.
In addition to the models in Figure~\ref{fig:zeta}, we also include
constraints from fitting only 2PCFs statistics ($w_p$ and $\xi_{0,2,4}$). The
left panel presents the best-fitting mean occupation functions. The mean
occupation functions of central galaxies can already be well constrained by
$w_p$, and the constraints from other models show only small variations. The
well-constrained mean occupation function of central galaxies explains why
the 3PCFs on large scales are similar for the different models. The
best-fitting mean occupation functions from $w_p$-only or from including
$\zeta$ are similar. Including all the 2PCFs and 3PCFs lead to a shallower
high-mass slope. However, the uncertainty in the high-mass slope is large, as
indicated by the shaded region (1$\sigma$) for the $w_p + \xi_{0,2,4} +
\zeta$ fit, and the satellite mean occupation functions from the different
models are in good agreement. In fact, the uncertainty in the high-mass slope
is even larger if we do not include the 3PCFs. The 3PCFs thus tighten the
constraints on the high-mass slope, as expected from the larger weight
towards higher halo mass from the intra-halo galaxy triplets. However, we do
not find substantial tightening for the high-mass slope, which results from
the lack of 3PCF measurements on scales below $1\mpchi$.

The agreement between the mean occupation functions constrained from the
$w_p$-only fit and the joint fit, together with the difference in the
corresponding model 3PCFs, suggests that galaxy velocity bias can improve the
interpretation of the data. The right panel of Figure~\ref{fig:hod} shows the
velocity bias constraints (contours of 95\% confidence level) for the three
relevant models.

The velocity bias constraints from $w_p+\xi_{0,2,4}$ (2PCF only statistics)
are the same as in G15. When replacing $\xi_{0,2,4}$ with the redshift-space
3PCF $\zeta$, the velocity bias constraints originate from $\zeta$. Indeed,
the red contour in the right panel of Figure~\ref{fig:hod} shows that the
3PCF data favours the existence of central galaxy velocity bias $\alpha_c$.
Note that in this figure, the $w_p+\zeta$ (no $v$ bias) model corresponds to
the point at $\alpha_c=0$ and $\alpha_s=1$. The value of $\alpha_c$ from
fitting $w_p+\zeta$ is consistent with that from the 2PCF constraints. The
effect of the central velocity bias can be inferred from the comparison
between the $w_p+\zeta$ (no $v$ bias) fit and the $w_p+\zeta$ fit with
velocity bias for the 3PCFs on large scales in Figure~\ref{fig:zeta}. The
model with no velocity bias shows an over-prediction of $\zeta$ for nearly
degenerate triangle configurations (i.e. $\theta \sim 0$ and $\theta \sim
\pi$). Consider the situation in an overdense region, which appears squashed
in redshift space along the line of sight because of large-scale infall. The
effect of the central velocity bias is to smear out the redshift-space
distribution of galaxies, making the distribution less squashed. A less
squashed distribution reduces the possibility of finding degenerate triangle
configurations, and thus lowers the amplitude of $\zeta$ for $\theta \sim 0$
and $\theta \sim \pi$, leading to a better fit.

In terms of the satellite velocity bias, the 3PCFs prefer to have satellite
galaxies moving faster than dark matter ($\alpha_s>1$). Such a constraint
comes mainly from the small-scale $\zeta$, where the Fingers-of-God (FoG)
effect in the redshift-space distribution of galaxies contributes. The
$\alpha_s>1$ satellite velocity bias, in combination with the central
velocity bias, enhances the FoG. This situation causes the galaxy
distribution more extended, and such a dilution reduces the possibility of
finding small-scale galaxy triplets, lowering the small-scale 3PCF $\zeta$.
This effect clearly explains the difference seen in the small-scale $\zeta$
in Figure~\ref{fig:zeta} between the $w_p+\zeta$ fits with and without
velocity bias.  The satellite velocity bias constraint (red contour) from the
3PCFs shows a tentative tension with that (blue contour) from the 2PCFs, even
though it is at a level of $<2\sigma$. If the tension is confirmed with more
accurate measurements, the model would need to be improved, e.g. by
introducing freedom in the spatial distribution profiles of satellites inside
haloes (see the tests in G15). With the current data, we conclude that
combining the measurements of the 2PCFs and the 3PCF significantly
strengthens the constraints on the galaxy velocity bias, as demonstrated by
the smaller black contour in the right panel of Figure~\ref{fig:hod}.

\section{Conclusion}
\label{sec:discussion}

In this paper, we measure the redshift-space 3PCFs for a volume-limited
sample of LRGs ($0.48<z<0.55$) in the SDSS-III BOSS CMASS DR11 data, and
perform HOD modelling of the 3PCFs and 2PCFs. Similar to the case with the
2PCFs (G15), the 3PCF measurements favours the existence of galaxy velocity
bias, with which we are able to reproduce the observed galaxy 3PCFs
remarkably well on all scales larger than $1\mpchi$.

By combining with the 2PCFs, the galaxy 3PCFs tighten the constraints on the
HOD parameters, because the three-point distribution is more sensitive to the
galaxy occupations in more massive haloes \citep{Kulkarni07}. Both the
redshift-space 2PCFs or 3PCFs can be used to constrain galaxy velocity bias.
Either of them leads to a consistent central galaxy velocity bias of around
$0.25$, i.e. on average the central galaxy is in motion with respect to the
core of its host halo, corresponding to a 1D velocity dispersion of
${\sim}79\kms$ for halo masses around $2\times 10^{13}\msun$ (see G15). The
redshift error (${\sim}30\kms$, \citealt{Bolton12}; already accounted for in
the model) alone is not enough to account for the central velocity bias. As
discussed in G15, such a motion is consistent with those inferred from other
observations, including the extrapolation from the measurements in galaxy
clusters \citep{Lauer14}. The satellite velocity distribution is consistent
with that of the dark matter from either constraint and from the joint one.

As seen from the covariance matrix and the fitting results, the 3PCFs on
small scales ($r<7\mpchi$) have stronger constraining power on the HOD
parameters (including the velocity bias parameters). Since we limit our
measurements of the 3PCF to scales larger than $1\mpchi$ to reduce the fibre
collision effects, we have only small improvements on constraining the
satellite occupation function (compared to those from the 2PCFs) and the
constraints to the phase-space distribution of satellite galaxies within
haloes are loose. The satellite velocity bias is degenerate with the spatial
distribution of satellites within dark matter haloes (see Figure 11 of G15).
The 3PCF is more sensitive to the satellite distribution profile, since it
probes the profile with triangles of different shapes. The tentative tension
of the satellite velocity bias constraints between using $\xi_{0,2,4}$ and
$\zeta$ (right panel of Figure~\ref{fig:hod}) may indicate the departure of
the spatial distribution of satellites from that of the dark matter.
Therefore, the small-scale 3PCF measurements may serve as a powerful way of
understanding the satellite galaxy distribution within haloes and may provide
constraints to the halo assembly bias effect \citep{Zentner14}. We plan to
explore such a possibility in our future work by using survey samples free of
fibre collision effects and by considering small triplet separations.

\section*{Acknowledgments}
We thank the anonymous referee for the helpful and constructive comments. ZZ
was partially supported by NSF grant AST-1208891. YPJ is supported by
973-project 2015CB857000, NSFC-11320101002, and Shanghai key laboratory grant
No. 11DZ2260700. CL acknowledges the support of NSFC (11173045, 11233005,
11325314, 11320101002) and the Strategic Priority Research Program ``The
Emergence of Cosmological Structures'' of CAS (Grant No. XDB09000000). We
gratefully acknowledge the use of computing resources at Shanghai
Astronomical Observatory, from the High Performance Computing Resource in the
Core Facility for Advanced Research Computing at Case Western Reserve
University, and from the Center for High Performance Computing at the
University of Utah.

Funding for SDSS-III has been provided by the Alfred P. Sloan Foundation, the
Participating Institutions, the National Science Foundation, and the U.S.
Department of Energy Office of Science. The SDSS-III web site is
http://www.sdss3.org/. SDSS-III is managed by the Astrophysical Research
Consortium for the Participating Institutions of the SDSS-III Collaboration
including the University of Arizona, the Brazilian Participation Group,
Brookhaven National Laboratory, University of Cambridge, Carnegie Mellon
University, University of Florida, the French Participation Group, the German
Participation Group, Harvard University, the Instituto de Astrofisica de
Canarias, the Michigan State/Notre Dame/JINA Participation Group, Johns
Hopkins University, Lawrence Berkeley National Laboratory, Max Planck
Institute for Astrophysics, Max Planck Institute for Extraterrestrial
Physics, New Mexico State University, New York University, Ohio State
University, Pennsylvania State University, University of Portsmouth,
Princeton University, the Spanish Participation Group, University of Tokyo,
University of Utah, Vanderbilt University, University of Virginia, University
of Washington, and Yale University.


\begin{thebibliography}{}
 \providecommand{\href}[2]{#2}
  \providecommand{\eprint}[1]{\href{http://arxiv.org/abs/#1}{arXiv:#1}}

\bibitem[Akaike(1974)]{Akaike74} Akaike, H.\ 1974, IEEE Transactions on
    Automatic Control, 19, 716

\bibitem[\protect\citeauthoryear{{Berlind} \& {Weinberg}}{{Berlind} \&
  {Weinberg}}{2002}]{Berlind02}
{Berlind} A.~A.,  {Weinberg} D.~H.,  2002, \apj, 575, 587

\bibitem[\protect\citeauthoryear{Bernardeau et
    al.}{2002}]{Bernardeau02} Bernardeau F., Colombi S., Gazta{\~n}aga
    E., Scoccimarro R., 2002, PhR, 367, 1

\bibitem[\protect\citeauthoryear{Bolton et al.}{2012}]{Bolton12} Bolton
    A.~S., et al., 2012, AJ, 144, 144

\bibitem[\protect\citeauthoryear{{Coil} et~al.}{{Coil} et~al.}{2006}]{Coil06}
    {Coil} A.~L. et~al.,  2006, \apj, 644, 671

\bibitem[\protect\citeauthoryear{Dawson et al.}{2013}]{Dawson13} Dawson
    K.~S., et al., 2013, AJ, 145, 10

\bibitem[\protect\citeauthoryear{Eisenstein et al.}{2011}]{Eisenstein11}
    Eisenstein D.~J., et al., 2011, AJ, 142, 72

\bibitem[\protect\citeauthoryear{{Gazta{\~n}aga}, {Norberg}, {Baugh} \&
  {Croton}}{{Gazta{\~n}aga} et~al.}{2005}]{Gaztanaga05}
{Gazta{\~n}aga} E.,  {Norberg} P.,  {Baugh} C.~M.,    {Croton} D.~J.,  2005,
  \mnras, 364, 620

\bibitem[\protect\citeauthoryear{{Gazta{\~n}aga} \&
  {Scoccimarro}}{{Gazta{\~n}aga} \& {Scoccimarro}}{2005}]{Gaztanaga05b}
{Gazta{\~n}aga} E.,  {Scoccimarro} R.,  2005, \mnras, 361, 824

\bibitem[\protect\citeauthoryear{{Gaztanaga} \& {Frieman}}{{Gaztanaga} \&
  {Frieman}}{1994}]{Gaztanaga94}
{Gaztanaga} E.,  {Frieman} J.~A.,  1994, \apjl, 437, L13

\bibitem[\protect\citeauthoryear{Gunn et al.}{2006}]{Gunn06} Gunn J.~E., et
    al., 2006, AJ, 131, 2332

\bibitem[\protect\citeauthoryear{{Guo} \& {Jing}}{{Guo} \&
  {Jing}}{2009}]{Guo09b}
{Guo} H.,  {Jing} Y.~P.,  2009, \apj, 698, 479

\bibitem[\protect\citeauthoryear{{Guo}, {Li}, {Jing} \& {B{\"o}rner}}{{Guo}
    et~al.}{2014c}]{Guo14b} {Guo} H.,  {Li} C.,  {Jing} Y.~P.,
    {B{\"o}rner} G.,  2014b, \apj, 780, 139

\bibitem[\protect\citeauthoryear{{Guo}, {Zehavi} \& {Zheng}}{{Guo}
  et~al.}{2012}]{Guo12}
{Guo} H.,  {Zehavi} I.,    {Zheng} Z.,  2012, \apj, 756, 127

\bibitem[\protect\citeauthoryear{{Guo} et~al.,}{{Guo} et~al.}{2013}]{Guo13}
    {Guo} H.  et~al., 2013, \apj, 767, 122

\bibitem[\protect\citeauthoryear{{Guo} et~al.,}{{Guo} et~al.}{2015}]{Guo15}
    {Guo} H.  et~al., 2015, \mnras, 446, 578 (G15)

\bibitem[\protect\citeauthoryear{{Guo} et~al.,}{{Guo} et~al.}{2014b}]{Guo14a}
    {Guo} H.  et~al., 2014a, \mnras, 441, 2398

\bibitem[\protect\citeauthoryear{{Jing} \& {B{\"o}rner}}{{Jing} \&
  {B{\"o}rner}}{2004}]{Jing04}
{Jing} Y.~P.,  {B{\"o}rner} G.,  2004, \apj, 607, 140

\bibitem[\protect\citeauthoryear{{Jing}, {Mo} \& {Boerner}}{{Jing}
  et~al.}{1998}]{Jing98}
{Jing} Y.~P.,  {Mo} H.~J.,    {Boerner} G.,  1998, \apj, 494, 1

\bibitem[Konishi \& Kitagawa(2007)]{Konishi07} Konishi, S., \& Kitagawa, G.\
    2007, Information Criteria and Statistical Modelin, Springer
    Science+Business Media, LLC, New York, NY USA

\bibitem[\protect\citeauthoryear{{Kulkarni}, {Nichol}, {Sheth}, {Seo},
  {Eisenstein} \& {Gray}}{{Kulkarni} et~al.}{2007}]{Kulkarni07}
{Kulkarni} G.~V. et~al.,  2007, \mnras, 378, 1196

\bibitem[\protect\citeauthoryear{Lauer et al.}{2014}]{Lauer14}
    Lauer T.~R., Postman M., Strauss M.~A., Graves G.~J., Chisari N.~E.,
    2014, ApJ, 797, 82

\bibitem[\protect\citeauthoryear{{Li}, {Jing}, {Kauffmann}, {B{\"o}rner},
  {White} \& {Cheng}}{{Li} et~al.}{2006}]{Li06}
{Li} C. et~al.,  2006, \mnras, 368, 37

\bibitem[\protect\citeauthoryear{Li et al.}{2012}]{Li12} Li C., Jing Y.~P.,
    Mao S., Han J., Peng Q., Yang X., Mo H.~J., van den Bosch F., 2012, ApJ,
    758, 50

\bibitem[\protect\citeauthoryear{Mar{\'{\i}}n et
    al.}{2013}]{Marin13} Mar{\'{\i}}n F.~A., et al., 2013, \mnras,
    432, 2654

\bibitem[\protect\citeauthoryear{{McBride} et~al.,}{{McBride}
  et~al.}{2011}]{McBride11b}
{McBride} C.~K. et~al., 2011, \apj, 726, 13

\bibitem[\protect\citeauthoryear{{Miyatake} et~al.,}{{Miyatake}
  et~al.}{2013}]{Miyatake13}
{Miyatake} H.  et~al., 2013, ApJ, preprint(\eprint{1311.1480})

\bibitem[\protect\citeauthoryear{{Pan} \& {Szapudi}}{{Pan} \&
  {Szapudi}}{2005}]{Pan05}
{Pan} J.,  {Szapudi} I.,  2005, \mnras, 362, 1363

\bibitem[\protect\citeauthoryear{{Peacock} \& {Smith}}{{Peacock} \&
  {Smith}}{2000}]{Peacock00}
{Peacock} J.~A.,  {Smith} R.~E.,  2000, \mnras, 318, 1144

\bibitem[\protect\citeauthoryear{{Prada}, {Klypin}, {Cuesta},
    {Betancort-Rijo}
  \& {Primack}}{{Prada} et~al.}{2012}]{Prada12}
{Prada} F. et~al.,  2012, \mnras, 423, 3018

\bibitem[\protect\citeauthoryear{{Reid} et~al.}{2014}]{Reid14} {Reid} B.~A.
    et~al., 2014, \mnras, 444, 476

\bibitem[\protect\citeauthoryear{{Scoccimarro}, {Sheth}, {Hui} \&
  {Jain}}{{Scoccimarro} et~al.}{2001}]{Scoccimarro01}
{Scoccimarro} R. et~al.,  2001, \apj, 546, 20

\bibitem[\protect\citeauthoryear{{Skibba} \& {Sheth}}{{Skibba} \&
  {Sheth}}{2009}]{Skibba09b}
{Skibba} R.~A.,  {Sheth} R.~K.,  2009, \mnras, 392, 1080

\bibitem[\protect\citeauthoryear{Smee et al.}{2013}]{Smee13} Smee S.~A., et
    al., 2013, AJ, 146, 32

\bibitem[\protect\citeauthoryear{{Smith}, {Sheth} \& {Scoccimarro}}{{Smith}
  et~al.}{2008}]{Smith08}
{Smith} R.~E.,  {Sheth} R.~K.,    {Scoccimarro} R.,  2008, \prd, 78, 023523

\bibitem[\protect\citeauthoryear{{Szapudi} \& {Szalay}}{{Szapudi} \&
  {Szalay}}{1998}]{Szapudi98}
{Szapudi} S.,  {Szalay} A.~S.,  1998, \apjl, 494, L41

\bibitem[\protect\citeauthoryear{Wang et al.}{2004}]{Wang04} {Wang} Y.
    et~al., 2004, MNRAS, 353, 287

\bibitem[\protect\citeauthoryear{Wang et al.}{2007}]{Wang07} {Wang} Y.
    et~al.,  2007, \apj, 664, 608

\bibitem[\protect\citeauthoryear{{Zehavi} et~al.,}{{Zehavi}
  et~al.}{2002}]{Zehavi02}
{Zehavi} I.  et~al., 2002, \apj, 571, 172

\bibitem[\protect\citeauthoryear{{Zehavi} et~al.,}{{Zehavi}
  et~al.}{2011}]{Zehavi11}
{Zehavi} I.  et~al., 2011, \apj, 736, 59

\bibitem[\protect\citeauthoryear{{Zehavi} et~al.,}{{Zehavi}
  et~al.}{2005}]{Zehavi05b}
{Zehavi} I.  et~al., 2005, \apj, 630, 1

\bibitem[\protect\citeauthoryear{Zentner, Hearin, \& van den
    Bosch}{2014}]{Zentner14} Zentner A.~R., Hearin A.~P., van den Bosch
    F.~C., 2014, MNRAS, 443, 3044


\bibitem[\protect\citeauthoryear{{Zheng}}{{Zheng}}{2004}]{Zheng04} {Zheng}
    Z.,  2004, \apj, 610, 61

\bibitem[\protect\citeauthoryear{{Zheng} et~al.,}{{Zheng}
  et~al.}{2005}]{Zheng05}
{Zheng} Z.  et~al., 2005, \apj, 633, 791

\bibitem[\protect\citeauthoryear{{Zheng}, {Coil} \& {Zehavi}}{{Zheng}
  et~al.}{2007}]{Zheng07}
{Zheng} Z.,  {Coil} A.~L.,    {Zehavi} I.,  2007, \apj, 667, 760

\bibitem[\protect\citeauthoryear{{Zheng}, {Zehavi}, {Eisenstein}, {Weinberg}
    \&
  {Jing}}{{Zheng} et~al.}{2009}]{Zheng09}
{Zheng} Z. et~al.,  2009, \apj, 707, 554

\end{thebibliography}
\end{document}